\documentclass[twocolumn]{aastex62}
\usepackage[T1]{fontenc}
\usepackage{ae,aecompl}
\usepackage{epsfig,graphics}	
\usepackage{amsmath}
\usepackage{amssymb}	

\usepackage{color}
\usepackage{multirow}
\usepackage{booktabs}
\usepackage{tabularx}
\usepackage{varioref}
\usepackage{ulem}

\usepackage{natbib}
\pdfoutput=1

 \definecolor{darkspringgreen}{rgb}{0.09, 0.45, 0.27}
 \definecolor{amber(sae/ece)}{rgb}{1.0, 0.49, 0.0}

 \def\lsim{\mathrel{\rlap{\lower3.5pt\hbox{\hskip0.5pt$\sim$}}
   \raise0.5pt\hbox{$<$}}}      
 \def\gsim{~\rlap{$>$}{\lower 1.0ex\hbox{$\sim$}}}

\newcommand{\nii}{\mbox{[N\,{\sc ii}\,]\,}}
\newcommand{\cii}{\mbox{[C\,{\sc ii}\,]\,}}

 \submitjournal{ApJ}

\shorttitle{[N\,{\sc ii}\,] 205$\mu$$\rm m$ Emission in ID 141}
\shortauthors{Cheng et al.}

\received{30/08/2019}
\revised{24/04/2020}
\accepted{28/05/2020}

\submitjournal{ApJ}

\begin{document}

\title{ALMA [N\,{\sc ii}\,] 205 $\mu${\MakeLowercase m} Imaging Spectroscopy of the Lensed Submillimeter galaxy ID 141 at redshift 4.24}

\correspondingauthor{Cheng Cheng}
\email{chengcheng@nao.cas.cn}

\author{Cheng Cheng}
\affil{Chinese Academy of Sciences$-$South American Center for Astronomy (CASSACA), National Astronomical Observatories, Chinese Academy of Sciences, 20A Datun Road, Chaoyang District, Beijing 100101, China}
\affil{China-Chile Joint Center for Astronomy, Camino El Observatorio 1515, Las Condes, Santiago, Chile}
\affil{Instituto de F\'isica y Astronom\'ia, Universidad de Valpara\'iso, Avda. Gran Breta\~na 1111, Valpara\'iso, Chile}

\author{Xiaoyue Cao}
\affil{National Astronomical Observatories, Chinese Academy of Sciences (NAOC), 20A Datun Road, Chaoyang District, Beijing 100012, China}
\affil{School of Astronomy and Space Science University of Chinese Academy of Science 19 A Yuquan Rd, Shijingshan District, Beijing, 100049, China}

\author{Nanyao Lu}
\affil{Chinese Academy of Sciences$-$South American Center for Astronomy (CASSACA), National Astronomical Observatories, Chinese Academy of Sciences, 20A Datun Road, Chaoyang District, Beijing 100101, China}
\affil{China-Chile Joint Center for Astronomy, Camino El Observatorio 1515, Las Condes, Santiago, Chile}
 
\author{Ran Li}
\affil{National Astronomical Observatories, Chinese Academy of Sciences (NAOC), 20A Datun Road, Chaoyang District, Beijing 100012, China}
\affil{School of Astronomy and Space Science University of Chinese Academy of Science 19 A Yuquan Rd, Shijingshan District, Beijing, 100049, China}

\author{Chentao Yang}
\affil{European Southern Observatory, Alonso de Córdova 3107, Casilla 19001, Vitacura, Santiago, Chile}

\author{Dimitra Rigopoulou}
\affil{Astrophysics Department, University of Oxford, Oxford OX1 3RH, UK}

\author{Vassilis Charmandaris}
\affil{Department of Physics, University of Crete, GR-71003 Heraklion, Greece}
\affil{Institute of Astrophysics, FORTH, GR-71110, Heraklion, Greece}

\author{Yu Gao}
\affil{Department of Astronomy, Xiamen University, Xiamen, Fujian 361005, China}
\affil{Purple Mountain Observatory \& Key Lab of Radio Astronomy, Chinese Academy of Sciences (CAS), Nanjing 210033, China}

\author{Cong Kevin Xu}
\affil{Chinese Academy of Sciences$-$South American Center for Astronomy (CASSACA), National Astronomical Observatories, Chinese Academy of Sciences, 20A Datun Road, Chaoyang District, Beijing 100101, China}
\affil{China-Chile Joint Center for Astronomy, Camino El Observatorio 1515, Las Condes, Santiago, Chile}

\author{Paul van der Werf}
\affil{Leiden Observatory, Leiden University, P.O. Box 9513, 2300 RA Leiden, The Netherlands}

\author{Tanio Diaz Santos}
\affil{Nucleo de Astronomia de la Facultad de Ingenieria, Universidad Diego Portales, Av. Ejercito Libertador 441, Santiago, Chile}
\affil{China-Chile Joint Center for Astronomy, Camino El Observatorio 1515, Las Condes, Santiago, Chile}
\affil{Institute of Astrophysics, FORTH, GR-71110, Heraklion, Greece}

\author{George C. Privon}
\affil{Department of Astronomy, University of Florida, 211 Bryant Space Sciences Center, Gainesville, 32611 FL, USA}

\author{Yinghe Zhao}
\affil{Yunnan Observatories, Chinese Academy of Sciences, Kun-ming 650011, People's Republic of China}
\affil{Key Laboratory for the Structure and Evolution of Celestial Objects, Chinese Academy of Sciences, Kunming 650011, People's Republic of China}
\affil{Center for Astronomical Mega-Science, CAS, 20A Datun Road, Chaoyang District, Beijing 100012, People's Republic of China}

\author{Tianwen Cao}
\affil{China-Chile Joint Center for Astronomy, Camino El Observatorio 1515, Las Condes, Santiago, Chile}

\author{Y. Sophia Dai}
\affil{Chinese Academy of Sciences$-$South American Center for Astronomy (CASSACA), National Astronomical Observatories, Chinese Academy of Sciences, 20A Datun Road, Chaoyang District, Beijing 100101, China}
\affil{China-Chile Joint Center for Astronomy, Camino El Observatorio 1515, Las Condes, Santiago, Chile}

\author{Jia-Sheng Huang}
\affil{Chinese Academy of Sciences$-$South American Center for Astronomy (CASSACA), National Astronomical Observatories, Chinese Academy of Sciences, 20A Datun Road, Chaoyang District, Beijing 100101, China}
\affil{China-Chile Joint Center for Astronomy, Camino El Observatorio 1515, Las Condes, Santiago, Chile}

\author{David Sanders}
\affil{Institute for Astronomy, University of Hawaii, 2680 Woodlawn Drive, Honolulu, HI 96822, USA}

\author{Chunxiang Wang}
\affil{National Astronomical Observatories, Chinese Academy of Sciences (NAOC), 20A Datun Road, Chaoyang District, Beijing 100012, China}
\affil{School of Astronomy and Space Science University of Chinese Academy of Science 19 A Yuquan Rd, Shijingshan District, Beijing, 100049, China}

\author{Zhong Wang}
\affil{China-Chile Joint Center for Astronomy, Camino El Observatorio 1515, Las Condes, Santiago, Chile}
\affil{Chinese Academy of Sciences$-$South American Center for Astronomy (CASSACA), National Astronomical Observatories, Chinese Academy of Sciences, 20A Datun Road, Chaoyang District, Beijing 100101, China}

\author{Lei Zhu}
\affil{China-Chile Joint Center for Astronomy, Camino El Observatorio 1515, Las Condes, Santiago, Chile}

\begin{abstract}
We present an Atacama Large Millimeter/submillimeter Array (ALMA) observation of the Submillimeter galaxy (SMG) ID 141 at z=4.24 in the \nii 205 $\mu$m line (hereafter \nii) and the underlying continuum at (rest-frame) 197.6 $\mu$m. Benefiting from lensing magnification by a galaxy pair at $z=0.595$, ID 141 is one of the brightest $z>4$ SMGs.  At the angular resolutions of $\sim1.2$ to 1.5\arcsec\ (1\arcsec $\sim$6.9 kpc), our observation clearly separates and moderately resolves the two lensed images in both continuum and line emission at a signal-to-noise ratio$>5$ . Our continuum-based lensing model implies an averaged amplification factor of $\sim5.8$ and reveals that the delensed continuum image has the S\'ersic index$\simeq$0.95 and the S\'ersic radius of $\sim$0.18\arcsec\ ($\sim$1.24 kpc). Furthermore, the reconstructed \nii velocity field in the source plane is dominated by a rotation component with a maximum velocity of $\sim$300 km/s at large radii, indicating a dark matter halo mass of $\sim$10$^{12}M_{\odot}$. This, together with the reconstructed velocity dispersion field being smooth and modest in value ($<$100 km/s) over much of the outer parts of the galaxy, favors the interpretation of ID 141 being a disk galaxy dynamically supported by rotation. The observed \nii/CO (7-6) and \nii/\cii 158$\mu$m line-luminosity ratios, which are consistent with the corresponding line ratio vs. far-infrared color correlation from local luminous infrared galaxies, imply a delensed star formation rate of (1.8$\pm 0.6)$$\times$$10^3M_\odot \rm yr^{-1}$ and provide an independent estimate of the size of the star-forming region $0.7^{+0.3}_{-0.3}$kpc in radius.
\end{abstract}

\section{Introduction} \label{sec:intro}

Star formation regulates the interstellar medium (ISM) and enriches the chemical composition of a galaxy; star formation is one of the most fundamental drivers of the evolution of a galaxy.  With the recent advances in technology on the ground and in space, more and more luminous star-forming galaxies at $z > 2$ have been identified in submillimeter (submm) bands \citep{Blain2002}. Submillimeter galaxies (SMGs) are among the infrared-brightest star-forming galaxies in the early universe \citep{Casey2014}.  However, their large distances and dusty nature make it difficult to discern their internal galactic structures at kiloparsec or subkiloparsec scales.  Consequently, how to effectively and comprehensively characterize their star formation rate (SFR), and determine the dominant star formation mode \citep[e.g., merger-induced star formation vs. clumpy star formation disks; ][]{Tacconi2006, Tacconi2008, Agertz2009, Dekel2009, Dave2010} remains an acute and yet challenging task.

In view of the unprecedented spectral line mapping capabilities provided by the recently commissioned Atacama Large Millimeter/submillimeter Array \citep[ALMA, ][]{2009IEEEP..97.1463W}, \citet{Lu2015} explored a new spectroscopic approach for simultaneously inferring the SFR, SFR surface density ($\Sigma_{\rm SFR}$), and some molecular gas properties of a distant galaxy by measuring only the fluxes of the CO\,(7$-$6) line (rest-frame 806.652 GHz or 372$\mu$m) and either the \nii\ line at 205$\mu$m\ (1461.134 GHz; hereafter as \nii) or the \cii\ line at 158$\mu$m (1900.56 GHz; hereafter as \cii).   For local luminous infrared galaxies (LIRGs; with an 8-1000 $\mu$m luminosity $10^{11}\,L_{\odot}< L_{\rm IR} < 10^{12}\,L_{\odot}$) and ultraluminous LIRGs (ULIRGs, $L_{\rm IR} > 10^{12}\,L_{\odot}$), the CO\,(7$-$6) line luminosity, $L_{\rm CO(7-6)}$, can be used to infer the SFR of a galaxy with a $\sim$30\% accuracy, irrespective of whether the galaxy hosts an active galactic nucleus \citep[AGN; ][]{Lu2014, Lu2015, 2017ApJS..230....1L, Zhao2016}. Furthermore, the steep anti-correlation between the \nii/CO\,(7$-$6) (or \cii/CO\,(7$-$6)) luminosity ratio and the rest-frame far-infrared (FIR) color, $C(60/100)$ ($\equiv$ $f_{\nu}(60\mu m)/f_{\nu}(100\mu m)$), can be used to estimate $C(60/100)$ or the dust temperature $T_{\rm dust}$ \citep{Lu2015}. $C(60/100)$ is in turn related to $\Sigma_{\rm SFR}$ \citep{Liu2015, Lutz2016}. Such an indirect approach to estimating $\Sigma_{\rm SFR}$ is useful at high redshifts, where it is often challenging to resolve a galaxy in the FIR/submm.

In addition, these lines are among the most luminous gas cooling lines that are widely used to probe different gas phases in galaxies. Furthermore, they probe different gas phases in galaxies. The CO (7$-$6) line traces the warm (excitation temperature $T_{\rm ex}$ = 150 K) and dense ($n_{\rm crit}\sim 105\ \rm cm^{-3}$) molecular gas that is in close proximity to the location of current or very recent SF activity. As shown in \citet{Lu2014, 2017ApJS..230....1L} the spectral line energy distribution (SLED) of (U)LIRGs is generally peaking around the CO (7$-$6) line. Although the \cii line is considered a primary tracer of photon-dominated regions \citep{1985ApJ...291..722T, 1985ApJ...291..747T}, it can also arise from ionized gas, since it only takes $\sim$11.3 eV to turn C into C+. On the other hand, the \nii line comes exclusively from ionized gas, and traces mainly diffuse, warm ISM due to its low critical density \citep[44 cm\(^{-3}\); ][]{2006ApJ...652L.125O}. In summary, these lines form a valuable set of extinction-free probes into the SF and gas properties in galaxies, especially high-$z$ objects.

Observations of these important gas cooling lines are still scarce for high-$z$ galaxies. For example, the \nii line has been detected in only a handful of objects at $z \gtrsim 3$, \citep[e.g., ][]{Combes2012, Nagao2012, Bethermin2016, Pavesi2016, Harrington2019, 2019arXiv190602293C} and very few of them have been spatially resolved \citep[e.g., ][]{Decarli2012, Ferkinhoff2015, Lu2017}. We have carried out an ALMA program to complete this line set on a small sample of SMGs between $4 < z < 5.5$ \citep{Lu2017, Lu2018, Zhao2020} to attempt to not only characterize their SF properties but also gain 
valuable insights into the physical conditions of their ISM.  

In this paper, we present our ALMA observation of ID 141 (R.A. = 14:24:13.9; decl. = +02:23:04; J2000) at $z$ = 4.24. With an on-source exposure time of only $\sim$ 5 minutes, we have detected and moderately resolved the \nii emission of this galaxy. Being weakly gravitationally amplified, ID 141 is one of the brightest galaxies discovered by the {\it Herschel} Astrophysical Terahertz Large Area Survey ($H$-ATLAS) project \citep{Eales2010, Bourne2016, Valiante2016}. The galaxy has been detected in continuum at 250, 350, 500 $\mu$m \citep{Eales2010}, 870 $\mu$m, 880 $\mu$m, and 1.2 mm \citep{Cox2011}. These continuum measurements gave an estimated $T_{\rm dust}$ $\sim 38$ K and $L_{\rm IR} \sim(8.5\pm0.3)\times 10^{13} \mu_{L}^{-1} L_{\odot}$ \citep{Cox2011}, where $\mu_{L}$ is the amplification factor. The CO ($4--3$) and H$_2$O($2_{11}-2_{02}$) (752 GHz) and H$_2$O$^{+}$ (746GHz) have also been detected with high signal-to-noise ratios \citep[S/Ns; ][]{Cox2011, Omont2013, Yang2016}. \citet{Bussmann2012} observed ID 141 with Keck AO in $Ks$ band (resolution: $\sim 0.1''$) and with Submillimeter Array (SMA) in 880 $\mu$m continuum (resolution: $\sim 0.5''$), and showed that the lensing involves a dry merger galaxy pair at $z=0.595$. 

The delensed SFR of ID 141 is about 2000 $M_\odot/$yr \citep{Cox2011, Bussmann2012}, which is above the SFR upper limit that is estimated from the observed gas mass in high-$z$ SMGs and the corresponding freefall time scale \citep{Karim2013}, and quite rare even in high-$z$ SMGs \citep{Barger2014, Cowie2018}. With the help of the lensing magnification by the foreground dark matter halo, we can spatially resolve the dust continuum, and reconstruct the gas velocity field. However, almost all previous ID 141 observations are either from single-dish telescopes with large beam size (e.g. Herschel, APEX or IRAM), or from interferometry arrays with low spatial resolution \citep[e.g., Plateau de Bure Interferometer (PdBI) observations with a spatial resolution of about 3$''$, see ][]{Cox2011}. The high-resolution SMA observations (angular resolution about 0.6$''$) only reveals the continuum morphology \citep{Bussmann2012}, but the spatially resolved emission line observations of ID 141 are still scarce. Previous studies show that both galaxy merger and secular evolution can lead to a high SFR at high-$z$ \citep{Tadaki2018}. To understand the origin of the high SFR in ID 141, we observe the \nii and FIR continuum with ALMA at a resolution of about 1$''$, and model the lensing image carefully to estimate the SFR region size, obtain a reliable magnification factor, and recover the gas velocity maps.

We describe our observation and data reduction in Sec. 2 and present the results in Sec. 3. In Sec. 4, we analyze the observed \nii line and continuum emission, fit the dust continuum data by a gravitational lensing model, and infer the dynamic structures of ID 141. Throughout the paper, we assume a cosmological model with $H_0 = 71\rm km/s/Mpc$, $\Omega_{\rm m}$ = 0.27, and $\Omega_{\Lambda}$ = 0.73 \citep{2007ApJS..170..377S}. At $z$ = 4.24, 1\arcsec corresponds to 6.91 kpc.

\begin{figure*}[!t]
    \centering
    \includegraphics[width=0.95\columnwidth]{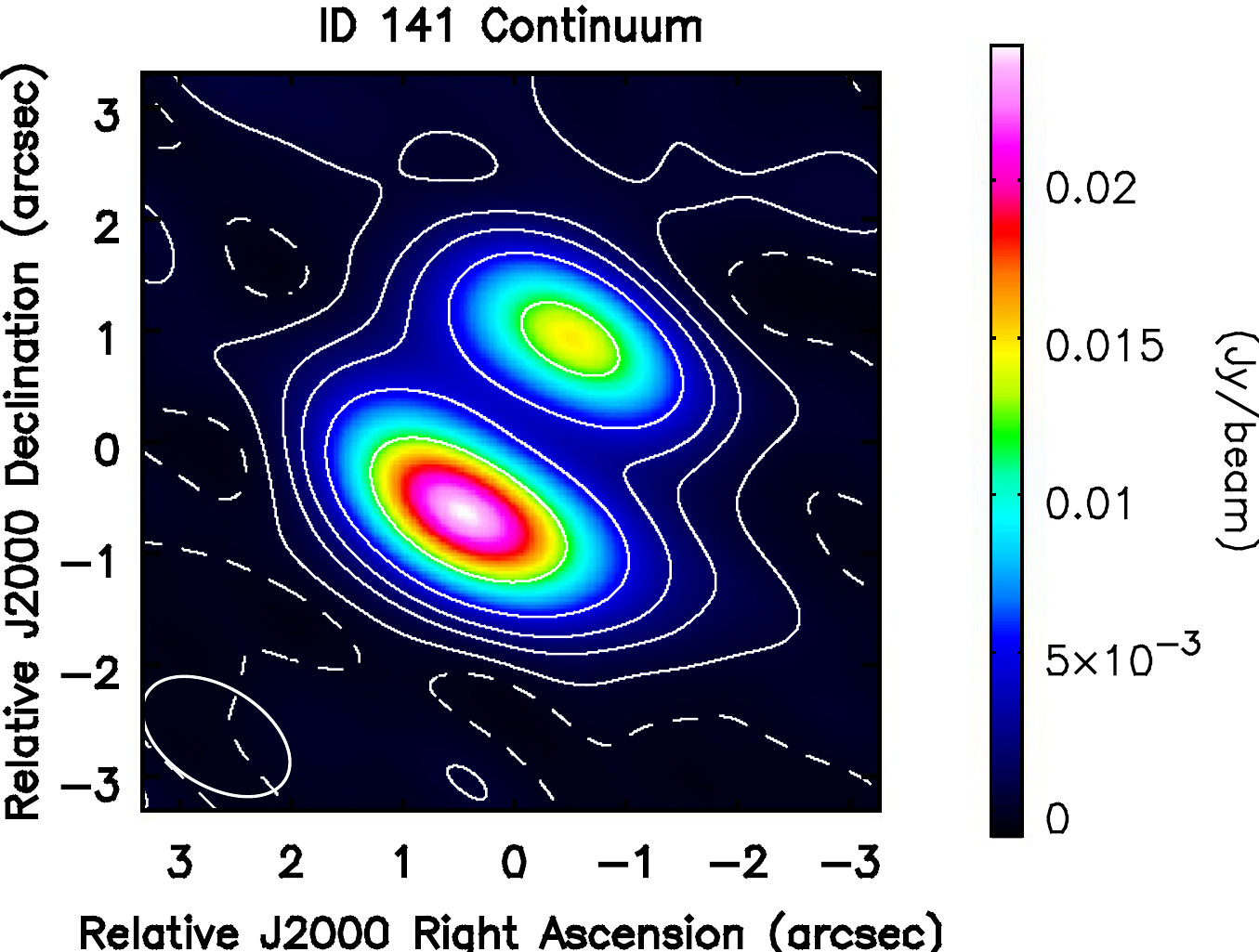}
    \includegraphics[width=0.95\columnwidth]{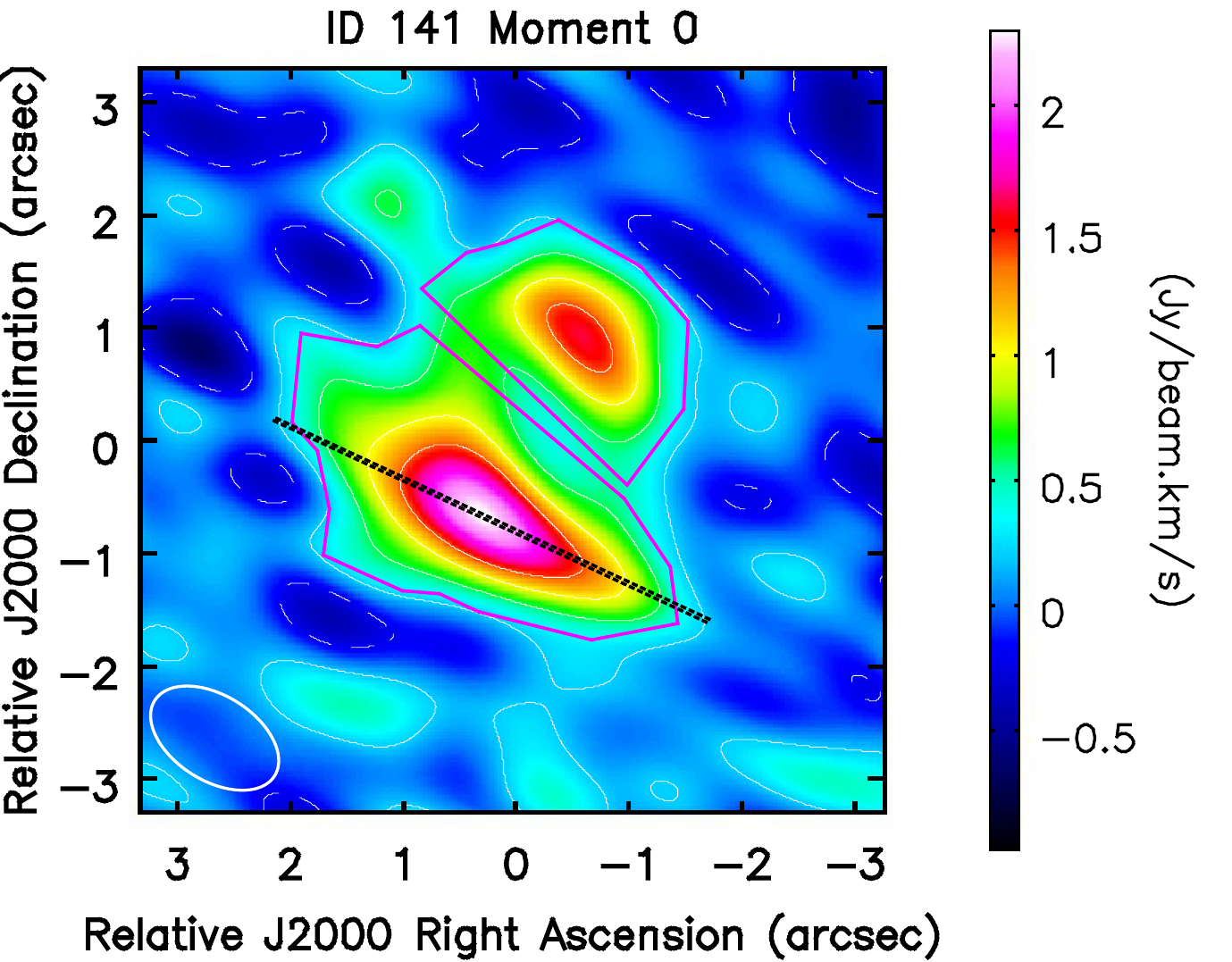}
    \includegraphics[width=0.95\columnwidth]{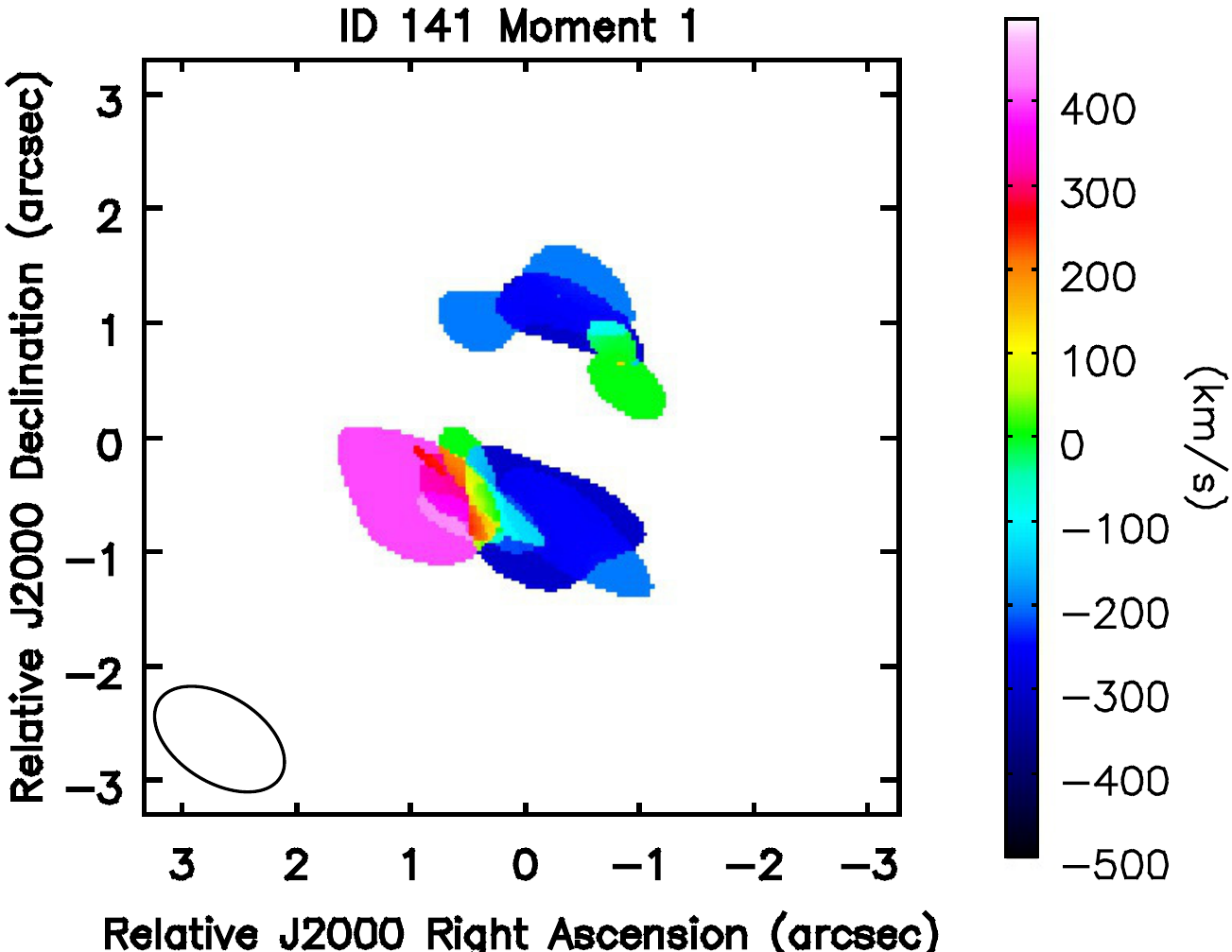}
    \includegraphics[width=0.95\columnwidth]{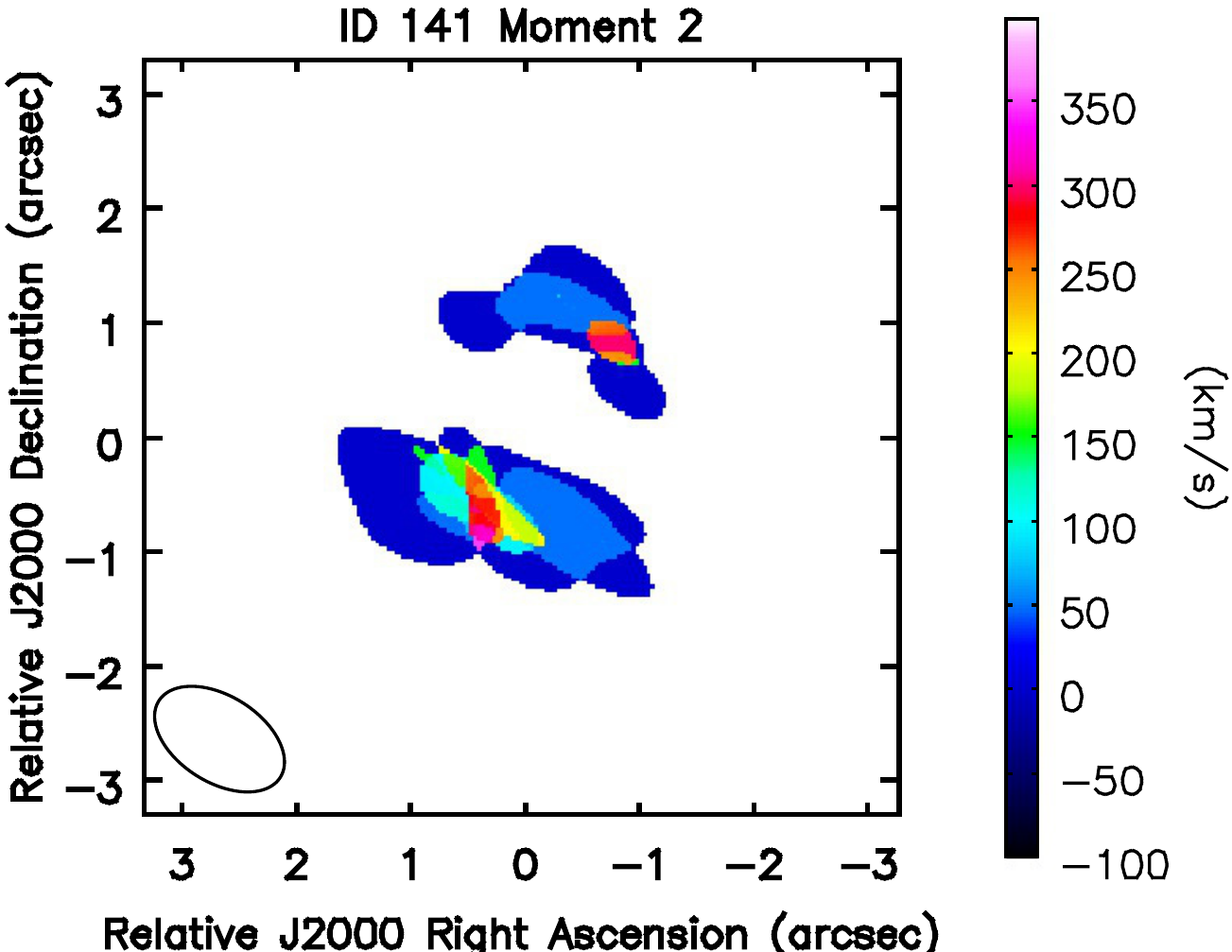}
    \caption{The continuum and the \nii moment maps for ID 141. The polygon regions in the moment 0 map show our two spectral extraction apertures. In the moment 1 map, the South component shows a clear rotation dominated features. The moment 2 map shows the velocity dispersion distribution, which for the South component is peaking at its nucleus, as expected. The rms of the moment 0 map is 0.2 Jy/beam km/s. The contours in the continuum image are at the [-1, 1, 5, 10, 20, 50] $\times$ rms level. The contours in the moment 0 show the [-1, 1, 3, 5, 7]$\times$ rms level. Dash contours stand for the -1 $\times$ rms. The position-velocity diagman along the direction of the black line marked on the moment-0 map is presented in Fig. \ref{pv}. The line is selected to pass through the long axis of the South component. 1'' corresponds to 6.9 kpc for this target.}
\label{ALMA_img}
\end{figure*}

\begin{figure}[ht!]
\centering
\includegraphics[width=0.45\textwidth]{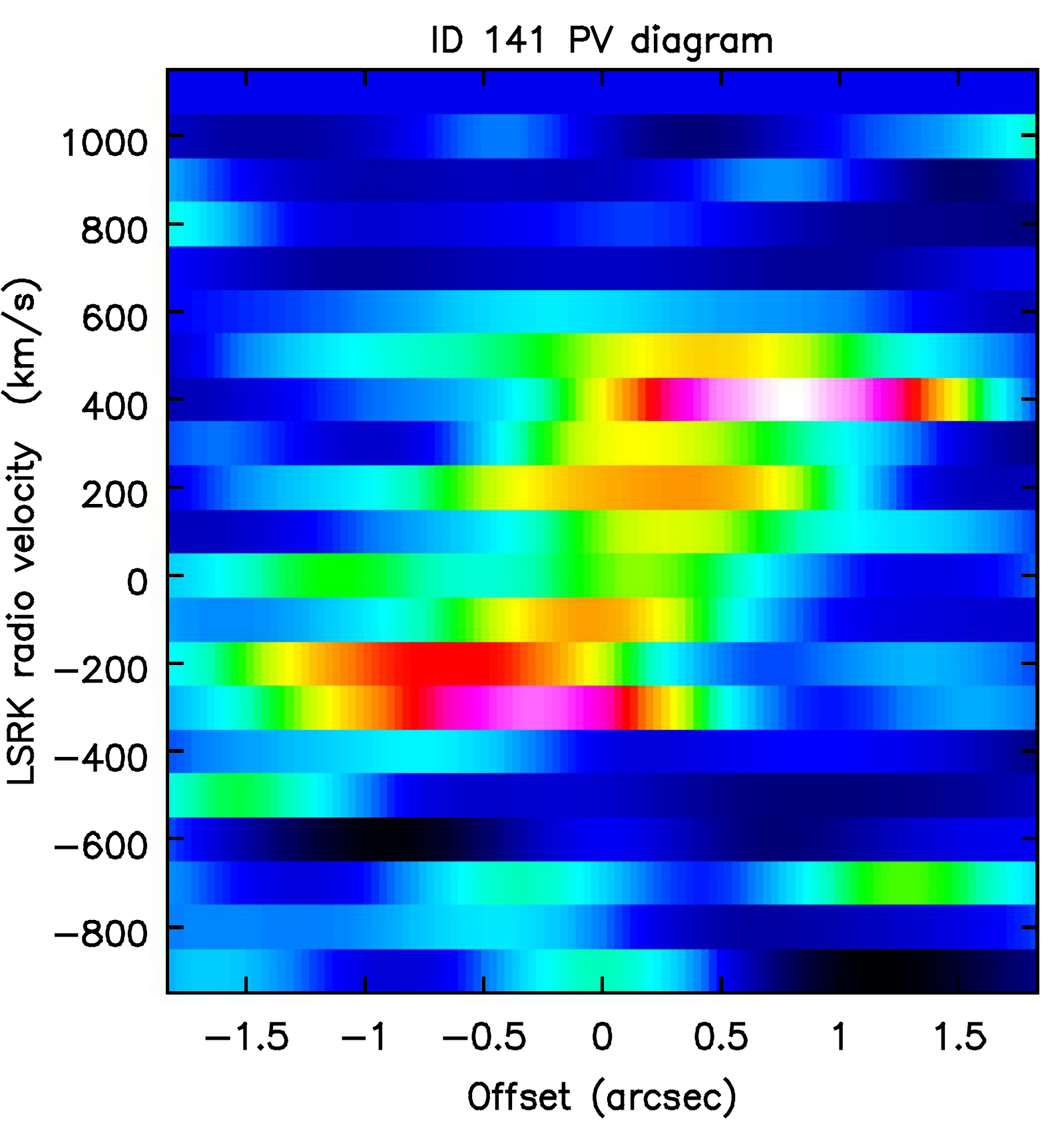}
\caption{
Position-velocity diagram along the black line in the moment 0 map. Limited by the angular and velocity resolution, we can only see the trend that the region with rotation velocity about 400 and -200 km/s is separated by 2 arcsec, which is about 2 times larger than the beam size.
}\label{pv}
\end{figure}

\begin{figure*}[ht!]
\centering
\includegraphics[width=0.45\textwidth]{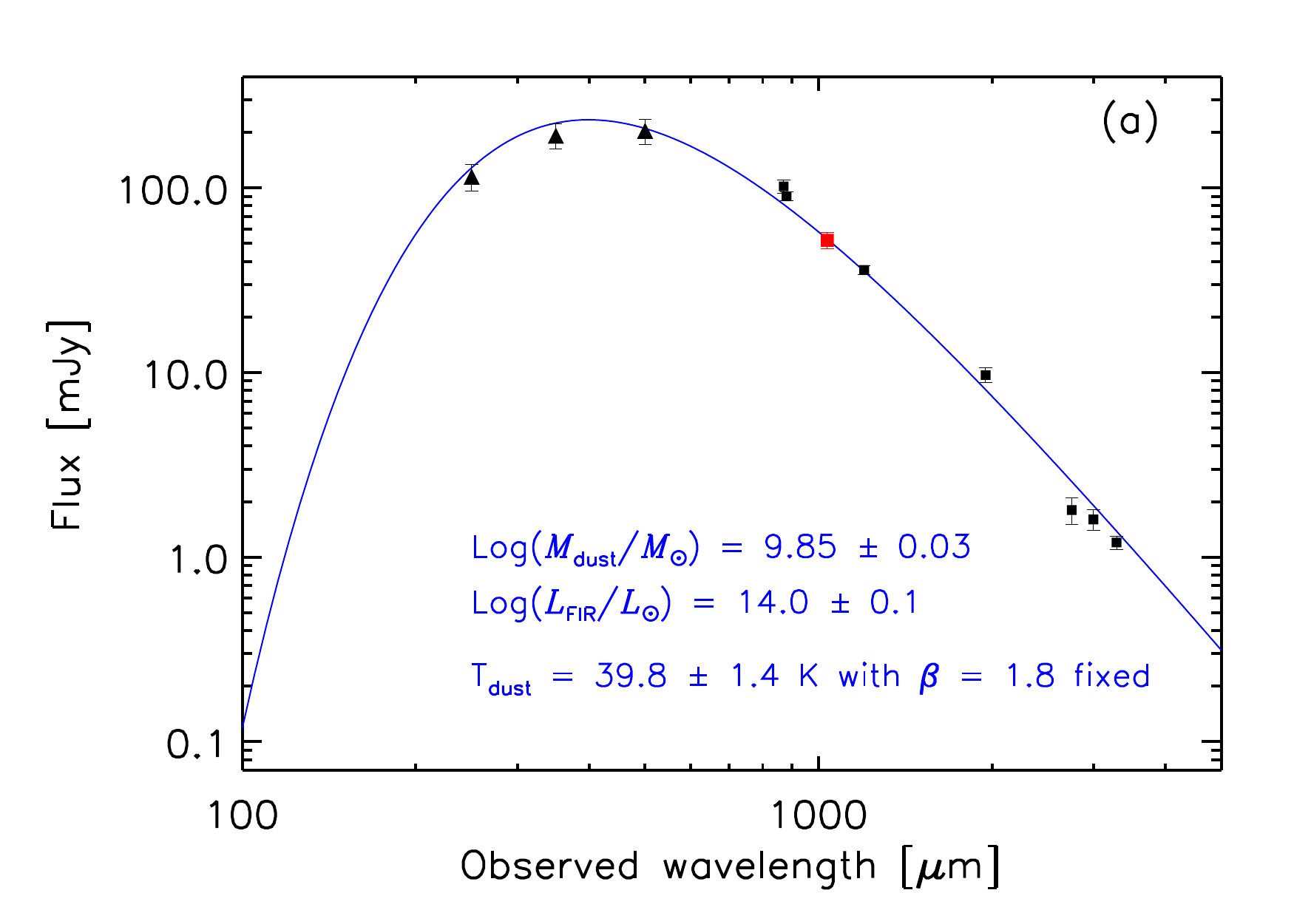}
\includegraphics[width=0.45\textwidth]{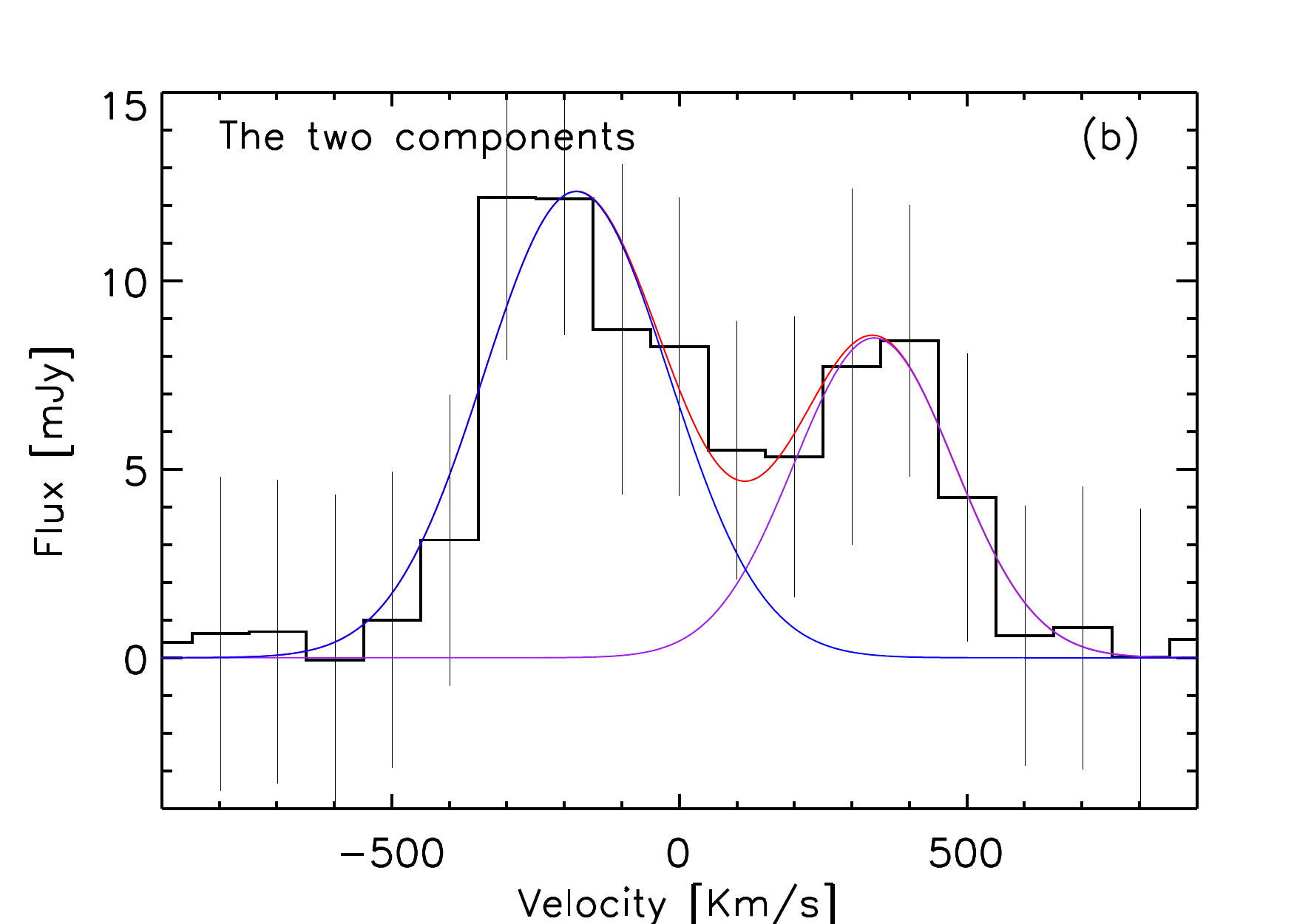}
\includegraphics[width=0.45\textwidth]{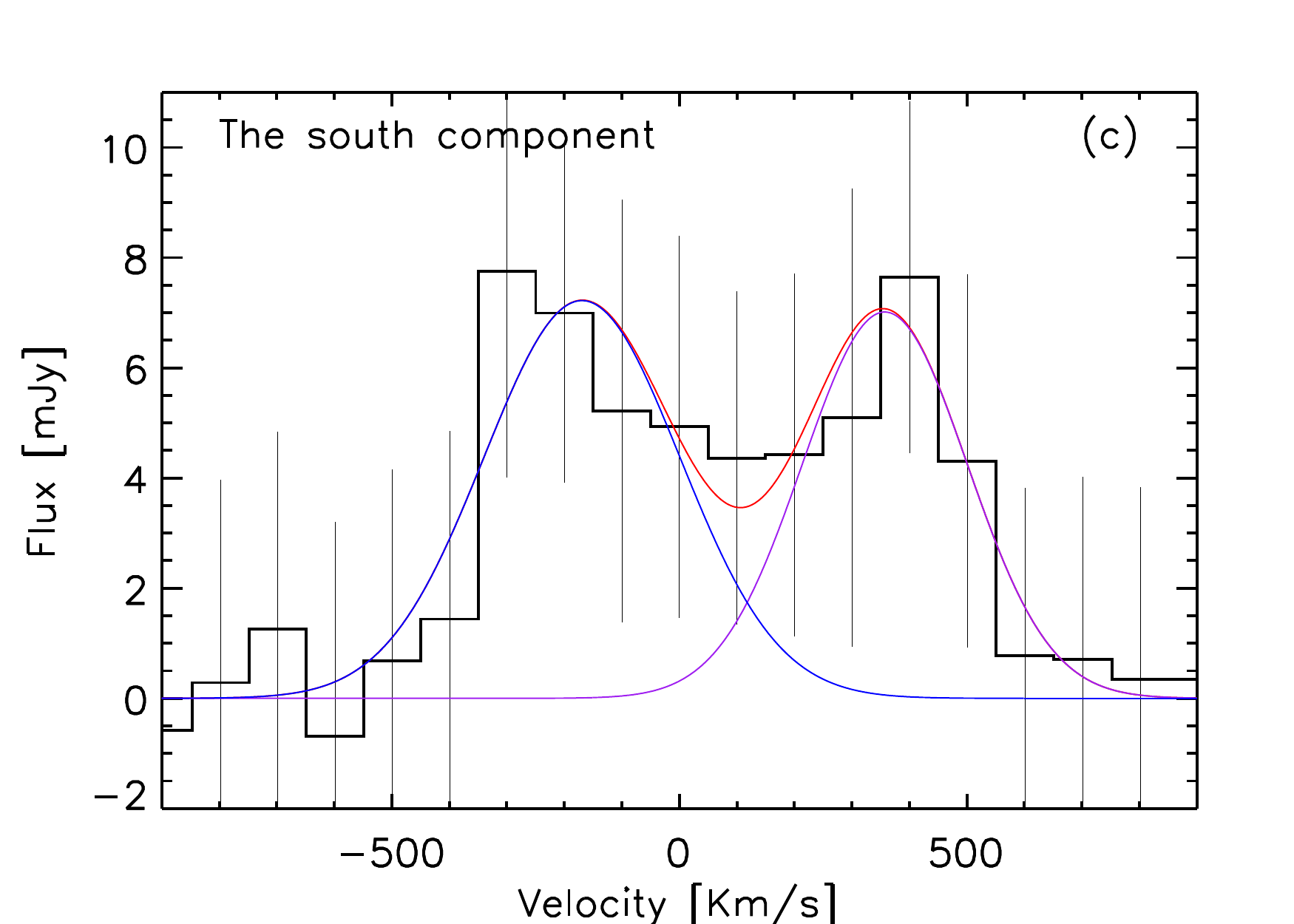}
\includegraphics[width=0.45\textwidth]{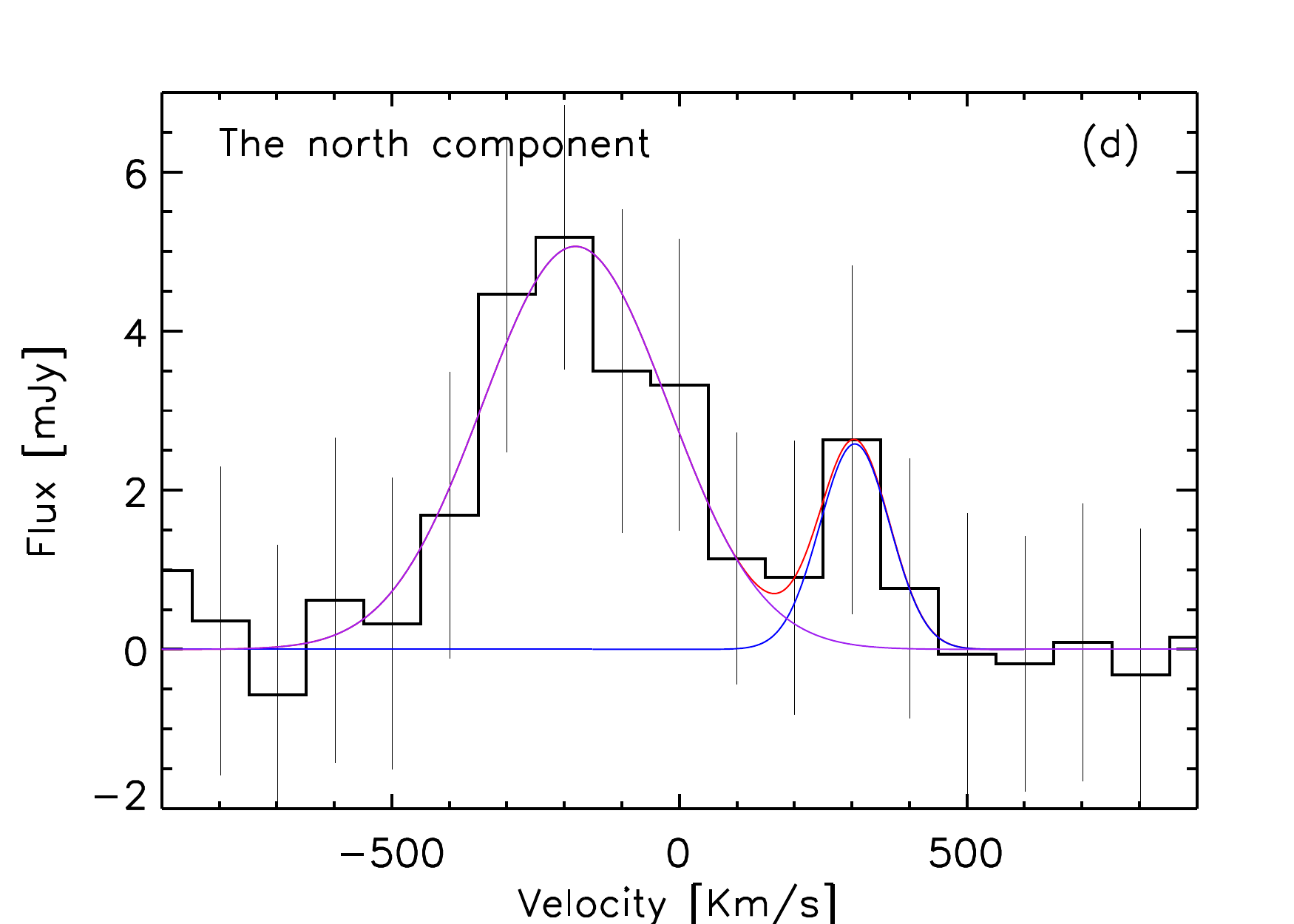}
\caption{Panel (a): 
FIR SED of ID 141 as observed on the plane of the sky.
The red dot is the new data from our ALMA observation. The upper triangles are the data from Herschel; the filled
squares are the data from the ground-based observations \citep{Cox2011}.
The ALMA flux are measured including both components shown in Fig. \ref{ALMA_img}. 
The dust emission is modelled by modified-blackbody function with a power law dust 
emissivity index $\beta = 1.8$, with the result shown by the solid curve. 
Panel (b): the total \nii spectrum extracted from the combined northern and 
southern polygon-shaped apertures shown in the moment 0 image in Fig .1,  
with the spectrum extracted from a single aperture shown in panel 
(c) (using the southern aperture) or panel (d) (the northern aperture).
The individual Gaussian components are shown by the curves in blue and purple, 
respectively; the combined Gaussian fit is shown by the curve in red.
The double peak function can fit the data well. Fitting results are listed in Table 1.
}\label{ALMA_sed}
\end{figure*}

\section{The ALMA observation and data reduction}

ID 141 is observed in ALMA Band 7. One of the four available spectral windows (with a bandwidth of 1.75 GHz ) was centered at the red-shifted \nii line at 277.4 GHz. The remaining three spectrum windows (SPWs) are used to measure the continuum around 277.7, 287.7, and 289.7 GHz, respectively. The on-target exposure time is 302.4 s. Each SPW has 128 channels with a channel width of 15.6 MHz and an effective spectrum resolution of 31.2 MHz. The observation utilizes 45 antennas and the baselines range from 15.1 to 331 m. The maximum recoverable angular scale associated with the smallest baseline used in our observations corresponds to about $9''$, much larger than the angular size of our target. The phase, bandpass and flux calibrations are based on the observations of J1359+0159, J1337-1257, and Callisto, respectively. The total observation time is 25 minutes. The phase center is at R.A. =14:24:13.98, decl. =02:23:03.50. (J2000)

The data reduction was carried out with the Common Astronomy Software Applications (CASA) 4.5.3 \citep{McMullin2007} and the final images are cleaned using the natural weighting only on the pixels with values $>3\sigma$, resulting in a synthesized beam size of $1.46'' \times 0.88''$ with a $\rm P.A. = 57.7^{\circ}$ for the continuum, and $1.25'' \times 0.8''$ with a $\rm P.A. = 57.8^{\circ}$ for the \nii line data. For the spectral data cube, the continuum is subtracted using the task ``uvcontsub'' with order$=1$. The final continuum image is an average of the three continuum SPWs and the RMS noise is 0.25 mJy beam$^{-1}$. The final \nii spectral cube has a velocity channel width of 100 km/s, and an RMS noise of about 0.6 mJy beam$^{-1}$ per velocity channel.


\begin{table}[h]
\centering 
\caption{Double gaussian fitting results of the \nii line in each component} 
\label{tab_label}
\begin{tabular}{ccccccccc} 
\hline
Component & line center (km/s) & FWHM (km/s) & flux (Jy km/s)\\ 
\hline
\hline
 south    &   -170 $\pm$ 81  &  401 $\pm$ 200  & 3.1 $\pm$ 1.3  \\
		  &    358 $\pm$ 76  &  336 $\pm$ 175  & 2.5 $\pm$ 1.2 \\
 north    &   -181 $\pm$ 49  &  381 $\pm$ 113  & 2.1 $\pm$ 0.5 \\
          &    305 $\pm$ 68  &  143 $\pm$ 109  & 0.4 $\pm$ 0.3 \\
south+north & -180 $\pm$ 51  &  378 $\pm$ 123  & 5.0 $\pm$ 1.4 \\
			&  339 $\pm$ 67  &  329 $\pm$ 161  & 2.9 $\pm$ 1.3 \\
\hline
\end{tabular}
\end{table}

\section{Results}

\subsection{Dust continuum and {\rm \nii} images}

The continuum and the \nii moment maps of ID 141 are shown in Fig. \ref{ALMA_img}. We denote the two clearly separated lensed components in the continuum and moment maps as the `north' and `south' components. For the continuum, the FWHM sizes along the long and short axes are respectively 1.84\arcsec\ and 0.96\arcsec\ for the south component and 1.78\arcsec\ and 1.00\arcsec\ for the north component. The corresponding FWHM dimensions for the \nii emission are 2.23\arcsec\ and 1.23\arcsec\ for the south component and 1.80\arcsec\ and 1.10\arcsec\ for the north component. These results show that the source is moderately resolved along its long axis in both continuum and \nii. We show the position-velocity diagram in Fig. \ref{pv}, measured along the black line shown in the \nii moment 0 map (Fig. \ref{ALMA_img}). The positions with 400 and -250 km/s are separated by more than 1 arcsec.

\subsection{{\rm \nii} Spectra}
We extract the \nii spectra from the two lensed image components by using polygon-shaped apertures, as outlined in red in the upper right panel of Fig. \ref{ALMA_img}. Panels (c) and (d) of Fig. \ref{ALMA_sed} are the spectra from the south and north components, respectively, and panel (b) of Fig. \ref{ALMA_sed} is the total spectrum from the two components combined. The flux uncertainty includes the flux calibration \citep[10\% of the flux, ][]{Fomalont2014} and the rms in the selected region. All spectra are fitted with a double-Gaussian function and the results are listed in Table 1. The total \nii flux is $7.9\pm 1.9$ Jy km/s. The FWHM of the total \nii line emission from this work is consistent, within the uncertainties, with the FWHM measurements from the CO and \cii 158 um lines given in \citet{Cox2011}, where the two lensed image components are not resolved. 

\subsection{Dust continuum}
\citet{Cox2011} provided the continuum fluxes at a few different wavelengths, but only for the two lensed components combined. Along with the continuum flux from this work, these are plotted in the upper left panel of Fig. \ref{ALMA_sed}. Our new ALMA continuum fluxes are measured from both components revealed from Figure \ref{ALMA_img}, with a resolution similar to the spectral energy distribution (SED) shown in \citet{Cox2011}. The total continuum flux measured by ALMA is 52.0 $\pm$ 5.2 mJy. We fit the FIR SED with a modified blackbody function \citep{Beelen2006}, with a fixed power-law index of the dust emissivity $\beta = 1.8$ \citep{Planck2011}. The resulting dust temperature is $\sim$40 K. 

We tested the cosmic microwave background (CMB) impact on our SED fitting, following \citet{Cunha2013}, and found an insignificant effect, largely due to the fact that the CMB is still much colder than the dust temperature of ID 141 at $z$ = 4.24.

\section{Analysis and Discussion}
\subsection{Lensing Modeling and Galaxy Intrinsic Properties}
To characterize the intrinsic morphology of the dust emission of ID 141, we need to trace the observed image from image-plane to source-plane via lens modeling. For the interferometer data, the incomplete sampling on the u-v plane will lead to spurious covariance features on the inverted dirty image. The ``Clean" method is usually taken to correct those pseudo signals before performing further analysis. However, the ``Clean" process will change the data in a way that is hard to quantify the uncertainties of cleaned-image and corresponding correlations. Poorly defined image uncertainties may bias the results of lensing modeling. Therefore, when comparing the data and model during lens-modeling, it is better to work on the u-v plane directly. This scenario has been the preferred modeling strategy when one deals with interferometric observations
\citep{Spilker16,Bussmann2012,Hezaveh13,Rybak15,Enia18}. 

In this work, we introduce our code -- tiny\_lens\footnote{\url{https://gitlab.com/cxylzlx/tiny\_lens}}. tiny\_lens is a light-weight galaxy-scale gravitational lens modeling tool that is originally designed for optical-band lenses. We add the visibility modeling capability to tiny\_lens, which is based on another open-source project -- Visilens\footnote{\url{https://github.com/jspilker/visilens}} \citep{Spilker16}. We summarize our modeling procedure here for completeness, and more technical detail can be found in \citet{Hezaveh13}.
\begin{enumerate}
\item Guess a set of possible parameter values associated with lens modeling, then generate an ``ideal image" based on them via ray-tracing.
\item Transform the ``ideal'' image from image-plane to uv-plane by Fourier transformation; this will give a visibility map on regular uv-grid. We then interpolate this ``regular'' visibility map to the ALMA uv-coordinates to get the ``model visibilities.''
\item The agreement between data and model visibilities is defined by the $\chi^2$ between them.  We use the nested sampling tool -- pymultinest\citep{Buchner2014,Feroz2009}, to iteratively ``guess'' the possible values of lens-modeling parameters, and sample the whole parameter space.
\end{enumerate}

As we have already mentioned before, previous works show that ID 141 is at redshift 4.24 lensed by two foreground galaxies located at redshift 0.595 \citep{Bussmann2012}. The ``reference model" in Bussmann's work adopted two Singular Isothermal Ellipsoids (SIE) for the mass distribution of the lens galaxies and a S\'ersic profile for the source light model. They also put a constraint on the mass ratio of two lens galaxies(2:1), since they found the secondary lens is significantly less luminous and hence likely to be less massive based on the Ks-band image of the Keck data.

In this paper, we model this system independently based on the ALMA 198 $\mu$m dust continuum. We take a similar modeling strategy as the ``reference model" in \citep{Bussmann2012}, which also consists of two SIEs for the lens and one S\'ersic profile for the source. However, we do not assume the mass ratio of two lens galaxies to be 2:1 in this work.

Our modeling results are shown in Table 2. In each row, from left to right, we present parameter names (units), prior types, prior ranges, median values and $1\sigma$ errors of the posterior probability distribution. The prior type can be uniform prior (type-0), Gaussian prior (type-1), or log-uniform prior (type-2). The corresponding meanings of prior ranges for each type prior are [lower bound, upper bound] (type-0 and type-2), [mean, standard deviation] (type-1). We report the center of main lens galaxy ($x_{L0}$, $y_{L0}$), the Einstein radius ($\theta_{E0}$), two ellipticity parameters ($e1_{L0}$, $e2_{L0}$)\footnote{In practice, working on ellipticities instead of position angle and axis ratio can improve the sampling efficiency, especially when the axis ratio is close to $\sim$1, of which the position angle effectively has no constraint.}. The definition of ellipticity can be found in \citet{Birrer15} and reproduced here,
\begin{equation}
    \Bigg(e1,e2 \Bigg) = \Bigg(\frac{1-q}{1+q}\cos(2\theta),\frac{1-q}{1+q}\sin(2\theta)\Bigg ).
\end{equation}
Here,  $e1$ and $e1$ are two ellipticity parameters, $q$ is axis ratio, and $\theta$ is the position angle. We fix the position offset between the secondary lens and the main lens to (-0.025\arcsec, -0.327\arcsec) during lens modeling, using the astrometry information provided by Keck image \citep{Bussmann2012}. The Einstein radius ($\theta_{E1}$), two ellipticity parameters ($e1_{L1}$, $e2_{L1}$) of secondary lens are also shown in Table 2. The brightness distribution of source galaxy are given by center ($x_{S}$ and $y_{S}$), Sersic index ($n_{S}$), two ellipticity parameters ($e1_{S}$ and $e2_{S}$), effective radius ($R_{e}$), and intensities at effective radius ($I_{e}$). We should note that the errors reported in Table 2 are purely statistical and do not include systematic errors. One of the predominant sources of systematical error in our case is we use a single over-simplified S\'ersic profile to represent the brightness distribution of source galaxies. We check the effect of this systematical error by the mock data test. We empirically find the single S\'ersic source assumption does introduce some systematics, but the lens parameter we inferred is still correct under the accuracy level of $\sim$10\%.

\begin{figure*}[ht!]
\includegraphics[width=0.99\textwidth]{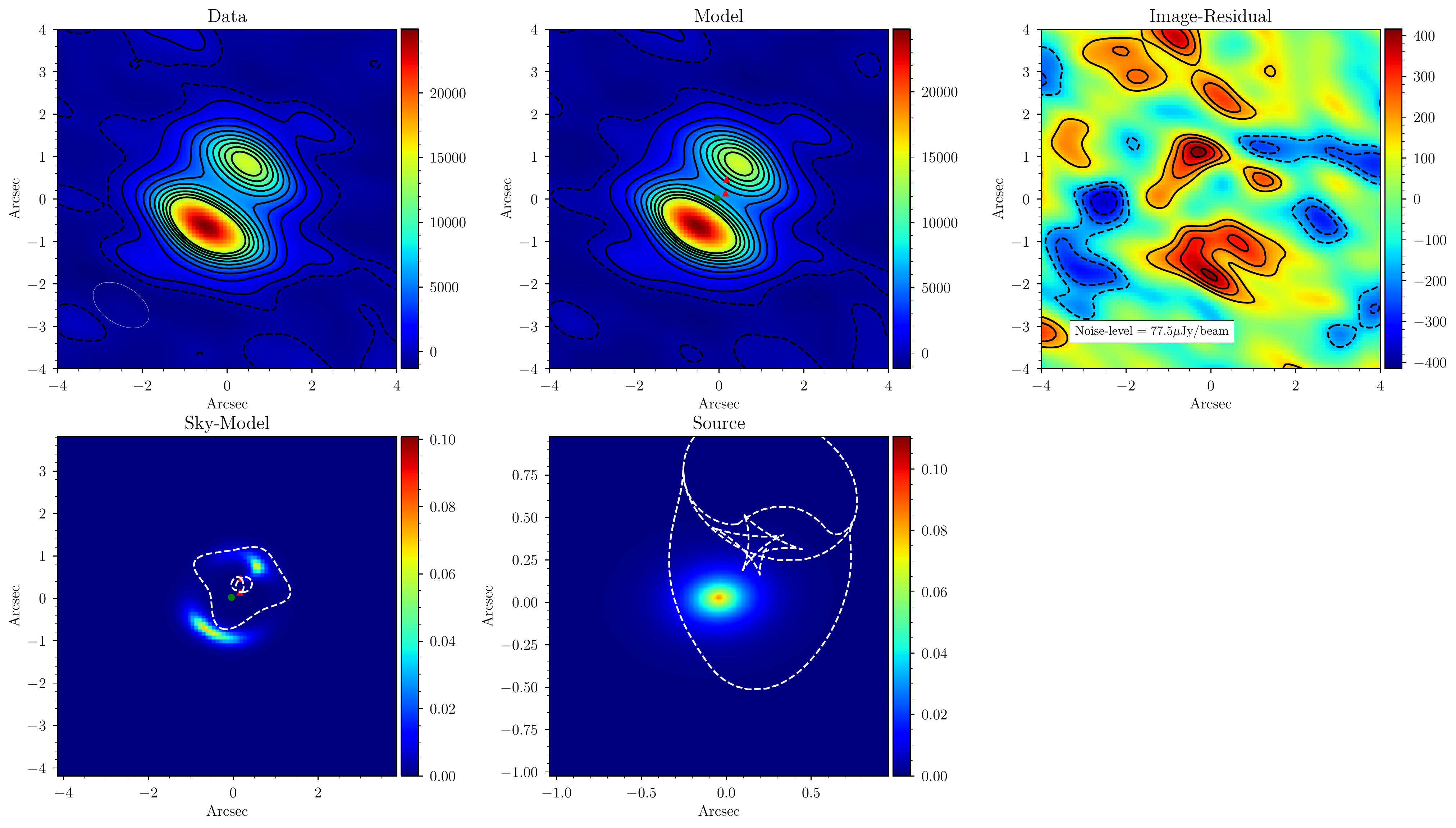}
\caption{
Top panels from left to right: the dirty image of the continuum data, model, and residual (data minus model) in unit of $\mu$Jy/beam. We show the beam by a white ellipse in the top-left panel. Contours in the `data' and `model' panel indicate -25, -5, 5, 25, 45, 65, 85, 105, 125, etc. times the $1\sigma$ RMS noise level, while Contours in the `residual' panel show -6, -5, -4, -3, -2, 2, 3, 4, 5, 6, etc. times the $1\sigma$ RMS noise level. The position of two lens galaxies and the source galaxy is displayed by the red-triangles and blue dot in the top-middle panel, respectively. The ``Noise-level" in the legend of the top-right panel shows the sum of visibility weights, i.e., $\sqrt{\sum_{i=1}^N{1/\sigma_i^2}}$, where the $\sigma_i$ represents the noise of each visibility. Bottom-right panel: the ideal image of the best-fit sky model in arbitrary units, the critical lines are shown by the black dashed line. Bottom-right panel: the best-fit source model in arbitrary unit, black dashed lines indicate the caustic lines.}
\label{lensing}
\end{figure*}

We visualize our best-fit results in Fig.\ref{lensing}, we find that our simplified lens model can already capture the main lensed feature in data. The best-fit model has $\chi^2 = 757972$ with 746981 degrees of freedom in the uv-plane, which corresponds to a reduced $\chi^2$ about 1.01. There is a $\sim$5 sigma residual feature on the northeast of the northern image. We anticipate those residuals are due to our simplified assumptions that source galaxies are represent by a single S\'ersic component. When higher quality data with better signal to noise ratio and u-v coverage are available in the future, it is possible to reveal the complex morphology of source galaxies using the more dedicated pixelized-model\citep{Rybak15,Hezaveh2016,Enia18,Dye2018}. However, such a task is beyond the scope of this work, we would like to leave it for future works. We note that the effective radius of the source galaxy in our results is 0.18\arcsec, which is significantly different from the results in \citet{Bussmann2012}. To further examine this difference, we fixed the source galaxy size to Bussmann's value (0.46\arcsec). Modeling results under this assumption decrease the Bayesian evidence by a factor of $\sim$500. Generally speaking, differences of more than ten might definitely rule out the model with the lower Bayesian evidence. Thus our results should be supported better by the current data. We also check this point independently based on the cleaned image, using another lens modeling tool--PyAutoLens \citep{Nightingale18}. We obtain a result that is similar to our tiny\_lens code.

\begin{figure}[ht!]
\centering
\includegraphics[width=0.5\textwidth]{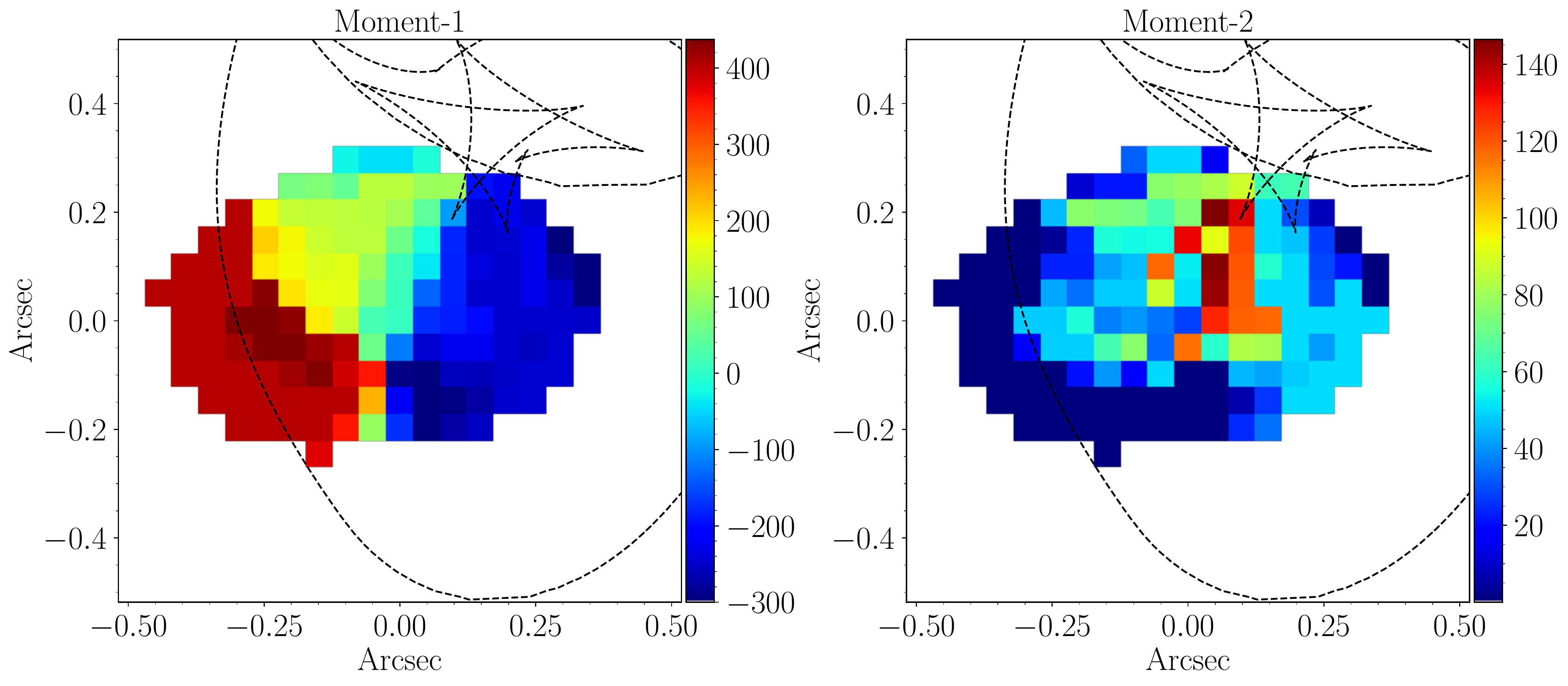}
\caption{
The \nii velocity and velocity dispersion map of ID 141 on the source plane. The map size is limited to the 0.36\arcsec, which is twice the half light radius of the source galaxy in intermediate-axis convention. We can see a clear rotation pattern in the velocity map. The velocity dispersion dominates the velocity field in the central region of ID 141 while the rotation dominates the outer regime. The pixel size in this figure are chosen as 0.052$\arcsec$. The caustics lines are marked by the black dashed line. $1\arcsec$ corresponds to 6.9 kpc at $z=4.24$.
}
\label{velocity-nii}
\end{figure}

Our lensing model result suggests that the ID 141 system is well described by a background S\'ersic source lensed by two foreground SIE lenses. The Einstein radii of the two foreground SIE lenses are 0.516\arcsec and 0.497\arcsec. These correspond to masses of $6.34\times10^{10}M_{\odot}$ and $5.88\times10^{10}M_{\odot}$, respectively. The axis ratio of the main lens is 0.71, which is more elliptical than the secondary lens (axis ratio:0.94). There is a mass-degeneracy between two lens galaxies: i.e., increasing the mass of one lens while decreasing the mass of another can result in a similar lensed image. This degeneracy implies that the total mass of two lens galaxies is better constrained than the individual lens mass given the current data. The S\'ersic index ($n_{S0}$) of the source galaxy is 0.95, which is between the 0.5 (corresponding to a Gaussian light profile) and 1.0 (corresponding to an exponential disk); this indicates the morphology of the source is close to an exponential disk or a combination of the typical Gaussian and exponential disk profile. A dust morphology comprised of a core plus a disk structure is also found in SMG G09v1.97 at $z$ = 3.63 \citep{Yang2020}. The half-light radius of the source is 0.18\arcsec, which is equivalent to 1.24 kpc in the physical units. 

As a natural telescope, a strong gravitational lensing system can be used to spatially resolve the velocity field of background source galaxies \citep{Stark08,Jones10,Dye15,Livermore15, Motta2018, Litke2019, Yang2019}. Following the discussions in \citet{Dye05}, `the source pixel size should be no smaller than Nyquist sampling of the PSF inverted to the source plane.' The beam size of our \nii emission data is $1.25\times0.8\arcsec$. Considering an average magnification of $\sim$5.8, to resolve the target in the source plane, the necessary of minimum pixel size is about $\sqrt{1.25\times0.8}/(2\times\sqrt{5.8}) = 0.21\arcsec$. Although this is only a rough estimation, recalling that the effective radius of our source galaxy is about $\sim$0.2\arcsec, we can conclude that ID 141 is only partially resolved in \nii emission for our current data. Since the spatial resolution and S/N ratio of the moment map derived from the \nii emission data are not very high, instead of using a sliced 3D data cube to reconstruct the source velocity field \citep{Dye15}, or modeling the source kinematics and lens-mass distribution simultaneously \citep{Rizzo18}, we just use the lens model derived from the continuum to trace the \nii moment map back to the source planes to that have multiple counterparts on the image plane; their values are obtained by stacking their image-plane values weighted by ﬂux. Although this method does not account for the beam smearing effect properly, it still offers a qualitative illustration of the kinematics of the source galaxy.

The resulting rotation and dispersion velocity fields are shown in Fig. \ref{velocity-nii}. The velocity field shows that ID 141 is rotation-dominated in the disk region and dispersion-dominated in the central part, which is commonly seen in the local disk galaxies. A few studies have cautioned that the apparent rotational feature in the first moment images of some high-$z$ galaxies may reflect a compact merger \citep{Simons2019, Yang2019}. The S\'ersic index of the compact galaxies is usually larger than 2, while the S\'ersic index value of ID 141 is 0.95, which is also commonly seen in disk galaxies. Therefore, we favor the conclusion that an orderly rotating disk is the source of the observed velocity field of ID 141. Higher spatial resolution ALMA observations will help us to reveal the gas instability and the origin of the high SFR density \citep{Tadaki2018}.

\begin{table}[h]
\centering 
\caption{Lensing model fitting results. We show the initial prior type and prior range for different lens modeling parameters. The median and $1\sigma$ errors of the posterior distribution are presented by the last two columns. The $1\sigma$ errors here are purely statistical errors; no systematical errors are included. All the positional information in this table are relative to the ALMA phase center (14:24:13.98, 02:23:03.50).} 
\label{lens_table}
\begin{tabular}{ccccccccc} 
\hline
name (unit) & prior type & prior range & median & $1\sigma$ \\ 
\hline
\hline
$x_{L0}$ (\arcsec) & 1 & [0.35 ,0.40] & 0.1796 & 0.0025 \\
$y_{L0}$ (\arcsec) & 1 & [0.75 ,0.40] & 0.4721 & 0.0135 \\
$\theta_{E0}$ (\arcsec) & 0 & [0.10 ,1.50] & 0.5162 & 0.0285 \\
$e1_{L0}$ & 0 & [-0.50 ,0.50] & -0.3722 & 0.0049 \\
$e2_{L0}$ & 0 & [-0.50 ,0.50] & 0.2932 & 0.0317 \\
$\theta_{E1}$ (\arcsec) & 0 & [0.10 ,1.50] & 0.4970 & 0.0295 \\
$e1_{L1}$ & 0 & [-0.50 ,0.50] & 0.2013 & 0.0310 \\
$e2_{L1}$ & 0 & [-0.50 ,0.50] & -0.3034 & 0.0112 \\
$x_{S}$ (\arcsec) & 0 & [-2.00 ,2.00] & -0.0430 & 0.0016 \\
$y_{S}$ (\arcsec) & 0 & [-2.00 ,2.00] & 0.0234 & 0.0039 \\
$I_{e}$ (Jy/beam) & 2 & [$10^{-9}$ ,0.10] & 0.0262 & 0.0007 \\
$R_{e}$ (\arcsec) & 0 & [0.01 ,1.00] & 0.1798 & 0.0016 \\
$n_{S}$ & 0 & [0.30 ,8.00] & 0.9539 & 0.0595 \\
$e1_{S}$ & 0 & [-0.50 ,0.50] & 0.1750 & 0.0125 \\
$e2_{S}$ & 0 & [-0.50 ,0.50] & 0.0276 & 0.0102 \\
\hline
\end{tabular}
\end{table}


The entire galaxy rotation curve spans only about three ALMA beams, which limits our ability to derive a detailed rotation curve. However, away from the center of the galaxy along the major axis, the typical velocity value is about 300 km/s. If the rotation curve is flat at large radii, the maximum rotational velocity should be around this 300 km/s value. The large velocity width observed also indicates a massive dark matter halo \citep[about \(4.7\times 10^{12}v_{300} M_\odot\), where the \(v_{300} = v_{\rm rot}/300\),][]{Ferrarese2002} at redshift 4.24, which is slightly higher than the typical high-z SMGs \citep[see Fig. 3 in ][]{Marrone2018}. Previous studies of the high-z SMGs \citep{Greve2005, Gullberg2015, Lu2017, Jones2017, Yang2017} have shown that the galaxy emission line FWHM or the maximum rotation velocity can be as high as 700 km/s, thus as one of the brightest high-z ULIRGs, ID 141 with such high rotation velocity may not be rare at high-z.

\subsection{Line ratio properties}

\begin{figure}[ht!]
\centering
\includegraphics[width=0.48\textwidth]{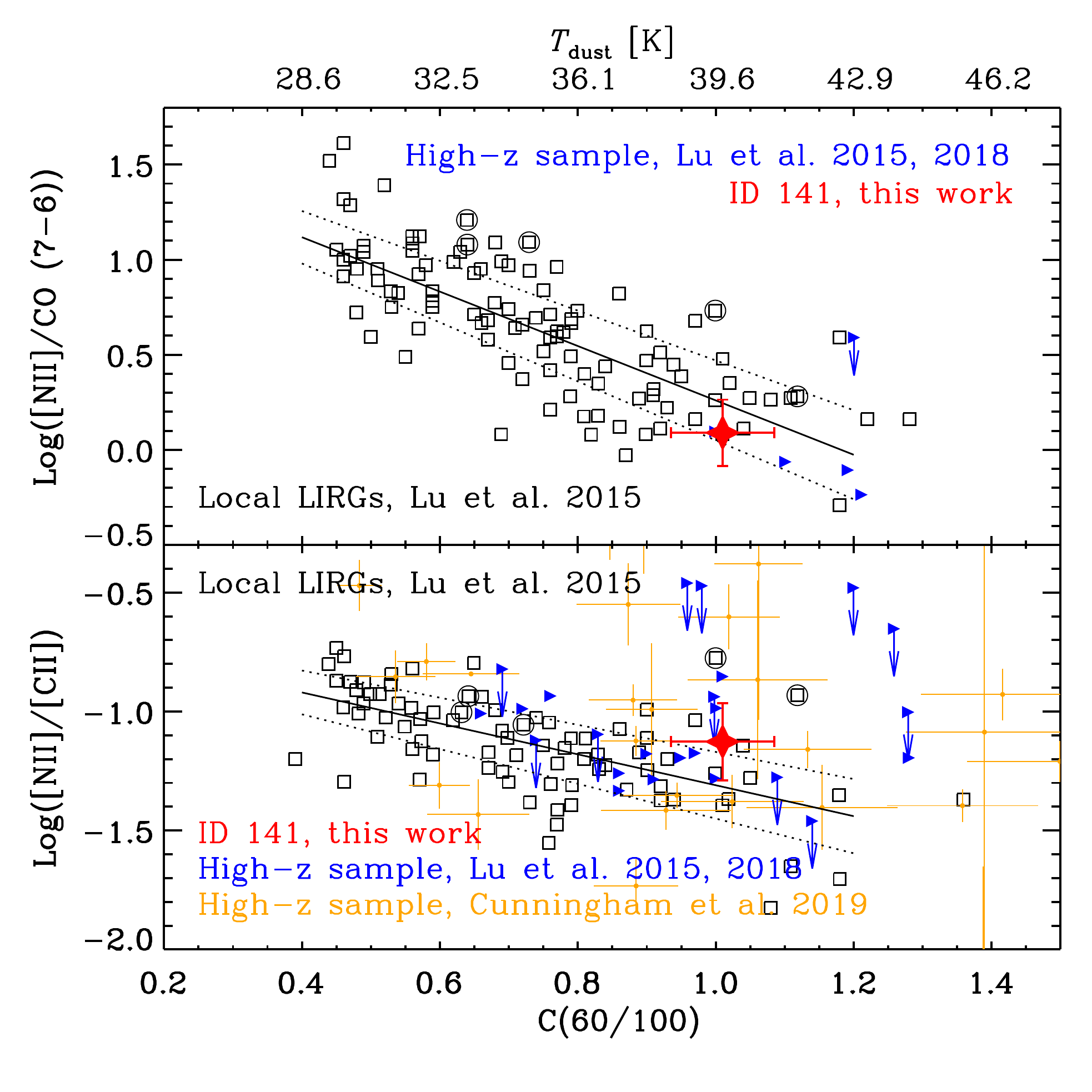}
\caption{
Comparisons between the \nii/CO (7-6), and \nii/\cii against the C(60/100) for the local LIRGs (open squares), high-z sample (blue triangles from \citet{Lu2018} and orange dots from \citet{2019arXiv190602293C}) and  ID 141 (the red star) presented in this work. The open squares with open circles represent the AGNs. The thick black line is the vertical least-squares fitting given in Eq. (4) and (5) in \citet{Lu2015}. The dot lines are the 1 $\sigma$ uncertainty of the fitting results. C(60/100) can be used as the dust temperature indicator \citep{Chanial2007, Santos2017}, so we also show the $T_{\rm dust}$ in the upper axis\textsuperscript{a}. For the local LIRG sample and the high-z sample from \citet{Lu2018}, the error bars are smaller than the sample scatter, so we omit them. The large scatter of the sample in \citet{2019arXiv190602293C} may caused by the presence of AGNs and the flux uncertainty in the sample.
}\label{line_ratio}
\small\textsuperscript{a} We derive the relation between the C(60/100) color and dust temperature by assuming a single temperature grey body SED model with the dust emissivity fixed to $\beta = 1.8$.
\end{figure}

\citet{Lu2015, Zhao2020} have shown that the \nii, \cii, and CO line ratios can be used to infer the dust temperature of galaxies at high-z with an accuracy of $\lesssim 4$K. Previous observations by \citet{Cox2011} measured the \cii flux at APEX and the CO (7-6) flux at PdBI with a beam size about $3.6''\times3''$. Our new \nii observation shows a flux of $7.9\pm 1.9$ Jy km/s, together with the \cii flux ($107\pm 17 $ Jy km/s) and CO (7-6) flux ($6.5\pm1.4$ Jy km/s) from \citet{Cox2011}, yielding line ratios of $\log$(\nii/\cii) = -1.12$\pm$0.16 and log(\nii/CO(7-6)) = 0.1$\pm$0.17. Fig. \ref{line_ratio} presents the line ratios \nii/\cii and \nii/CO (7-6) as a function of the flux ratio between the rest-frame 60 and 100$\mu$m, C(60/100). We include previous results for the local LIRGs (open squares) and high-z targets \citep[filled triangles, from ][]{Lu2015, Lu2018}. The thick lines in Fig. \ref{line_ratio} are the linear fitting results of the local LIRGs \citep[Eq. 3, 4 in ][where the AGNs are excluded in the fitting process]{Lu2015}, while the dot lines show the 1$\sigma$ uncertainty. Results of ID 141 line ratios offset the correlation at 1$\sigma$, so we conclude that the correlations between the line ratios and C(60/100) for local (U)LIRGs, as discussed in \citet{Lu2015} and \citet{Lu2018}, still hold, indicating a valid method to diagnose the high-z galaxy FIR color C(60/100) with 1$\sigma$ uncertainty. 

The line ratio can also help us to probe the gas properties such as the origin of the \cii and the metallicity. The \nii/\cii of ID 141 is $\sim$0.076, which is much lower than the typical value of HII regions \citep[\nii/\cii about 0.5, see][]{Decarli2014, Bethermin2016},  suggesting that most of the \cii flux may originate from the neutral gas, where the \nii is absent \citep{Decarli2014, Croxall2017, Sutter2019, 2019arXiv190602293C}. 

On the other hand, based on theoretical models of the metallicity effects to the line emission \citep{Kewley2002, Nagao2012, Bethermin2016, Pereira2017}, the low \nii/\cii value of ID 141 may also suggest a sub-solar metallicity of this high-z SMGs \citep{Nagao2012, Croxall2017}.

\subsection{Star Formation Properties}

The FIR SED fitting results in a dust temperature of about 40 K, which is warm and consistent with the high SFR of SMGs \citep{Magnelli2012}. We derive the $L_{\rm IR, obs}^{8-1000\mu \rm m} = 9.9 \pm 2.3\times 10^{13} \mu_L^{-1} L_\odot = 1.7 \pm 0.4 \times 10^{13} L_\odot$, where the lensing magnification factor $\mu_L = 5.8$. The SFR can be derived from the IR luminosity: SFR=$1.08\times 10^{-10} L_{\rm IR}^{8-1000 {\rm \mu m}}/L_\odot\,\,M_{\odot}/{\rm yr}$ \citep[][]{Kennicutt1998} or from the CO (7--6) luminosity: $8.18\times 10^{-6}L_{\rm CO \, (7-6)}/L_{\odot}$ \citep[][]{Lu2015}. Both formulae assume a Chabrier initial mass function \citep{Chabrier2003}, and find SFR$_{\rm FIR} = 1843 \pm 424 M_{\odot}/\rm yr$ and SFR$_{\rm CO(7-6)} \simeq 2256 \pm 423 M_{\odot}/\rm yr$, after the correction for the magnified factor $\mu_L = 5.8$. These SFR estimates are consistent with each other given the fact that the relative accuracy is $\sim$30\% between the two SFR estimators \citep{Zhao2020}.

From the greybody fit of the FIR SED, we derive that C(60/100) is about 1.01, corresponding to a $\Sigma_{\rm SFR} = 530\pm 210 M_{\odot}/\rm yr/kpc^2$ \citep{Liu2015, Lutz2016}. The half-light radius, $R_{\rm SF}$, of the star-forming region is given by $\Sigma_{\rm SFR} = {0.5\times\rm SFR} / (\pi \times R_{\rm SF}^2)$, resulting in $R_{\rm SF}$ = 0.74 $\pm$ 0.30 kpc. Here we use the SFR from a FIR SED fitting and take the difference between SFR$_{\rm FIR}$ and SFR$_{\rm CO(7-6)}$ into the uncertainty. This SFR surface density-weighted radius is comparable to the half-light radius (1.24 kpc) from our lensing model. Previous high-resolution FIR continuum observations revealed a positive correlation between the FIR half-light radius and the FIR luminosity \citep{Fujimoto2017} for the SMGs with $12\lesssim\log(L_{\rm FIR}/L_\odot)\lesssim 13$. If we apply the FIR size-luminosity relation to ID 141, we would obtain an FIR radius of about 2 kpc for $L_{\rm FIR} = 1.7\times 10^{13}L_\odot$, which is two times larger than the FIR radius of ID 141. So ID 141 may have a higher SFR surface density than the galaxies with similar FIR luminosity. Nevertheless, since our result is estimated from the ALMA observation with 1'' beam, additional higher spatial resolution data would help reveal more detailed star formation structures.

\section{Conclusion}

We present our recent ALMA band-7 observation of the $H$-ATLAS selected SMG: ID 141 in the \nii and continuum at rest-frame 197.6 $\mu$m at a spatial resolution of $1.2''$ and $1.5''$ ($1''$ $\sim$ 6.9 kpc). Taking advantage of the gravitational lens, our new ALMA observation helps us to moderately resolve the FIR continuum and dynamical structures. Our continuum-based lensing model fitting result reveals the FIR continuum has a S\'ersic index of about 0.95 and an effective radius of $\sim 0.18''$. The FIR size is about two times smaller than the typical SMG with the ID 141 intrinsic FIR luminosity being 1.7$\times 10^{13}L_\odot$. We further reconstruct the \nii velocity field on the source plane and find a rotation-dominated dynamical structure. The morphology and dynamics on the source plane suggest a high-$z$ disk SMG with a fast rotational velocity (about 300 km/s), indicating a dark matter halo mass of the order of $5\times 10^{12}M_\odot$. We develop our lensing model fitting code, which can apply a non-informative prior to our lens-modeling parameters, and shows good potential to study the gas distribution and gas gravitational instability \citep[e.g. Toomre parameter Q, ][]{1964ApJ...139.1217T} of ID 141 with higher-resolution data.

The observed line ratios (\nii/CO (7--6) and \nii/\cii) and the FIR color C(60/100) of ID 141 are consistent with the previous line ratio versus the FIR color correlation \citep{Lu2018}. Our method to estimate the high-$z$ galaxy dust temperature with the line-ratio measurements is valid for ID 141  with about 1 $\sigma$ uncertainty. Moreover, the \nii/\cii value suggests that most of the \cii line originates from the neutral gas and the metallicity of ID 141 is lower than $Z_\odot$. 

Together with the previous empirical relation between the star formation surface density and the FIR color, and the SFR derived from the CO (7--6) and FIR flux, we estimate a star formation radius of about 1 kpc, which is consistent with the size derived from our lensing model.

\section*{Acknowledgements}
We thank the referee for carefully reading and for patiently providing constructive comments that helped us to improve the quality of this paper. 
CC appreciates useful comments from Dr. gustavo Orellana Gonzalez and Alejandra Munoz Arancibia. 
This work is supported in part by the National Key R\&D Program of China grant 2017YFA0402704, 
the NSFC grant \#11673028, and the Chinese Academy of Sciences (CAS) through a grant to South 
America Center for Astronomy (CASSACA) in Santiago, Chile.
C.Y. was supported by an ESO Fellowship. GCP acknowledges support from the University of Florida.
DR aknowledges support from  UKRI grant ST/S000488/1.
C.C. is supported by the National Natural Science Foundation of China, No. 11803044,
and supported by the Young Researcher Grant of National Astronomical Observatories, Chinese Academy of Science. Y.G. research is supported by National Key Basic Research and Development Program of China (grant No. 2017YFA0402700), National Natural Science Foundation of China (grant Nos. 11861131007, 11420101002), and Chinese Academy of Sciences Key Research Program of Frontier Sciences (grant No. QYZDJSSW-SLH008). T.D-S. acknowledges support from the CASSACA and CONICYT fund CAS-CONICYT Call 2018. Y.Z. is supported by NSFC \#11673057.
J.H. is supported by NSFC \#11933003. RL acknowledge the support from NSFC (No, 11773032), and support from Nebula Talent Program of NAOC. This paper makes use of the following ALMA data: ADS/JAO.ALMA\#2015.1.00388.S. ALMA is a partnership of ESO (representing its member states), NSF (USA) and NINS (Japan), together with NRC (Canada) and NSC and ASIAA (Taiwan) and KASI (Republic of Korea), in cooperation with the Republic of Chile. The Joint ALMA Observatory is operated by ESO, AUI/NRAO and NAOJ. The National Radio Astronomy Observatory is a facility of the National Science Foundation operated under cooperative agreement by Associated Universities, Inc.



\end{document}